\documentclass[runningheads]{llncs}
\usepackage{graphicx,booktabs,array}

\begin{document}

\title{Conditional Score-Based Diffusion Model for Cortical Thickness Trajectory Prediction}

\author{Qing Xiao\inst{1,2\star}\and
Siyeop Yoon\inst{2\star}\and
Hui Ren\inst{2}\and
Matthew Tivnan\inst{2}\and
Lichao Sun\inst{3}\and
Quanzheng Li\inst{2}\and
Tianming Liu\inst{4}\and
Yu Zhang\inst{1\dag}\and 
Xiang Li\inst{2\dag}}

\authorrunning{Q. Xiao et al.}
\institute{School of Biomedical Engineering, Southern Medical University, Guangzhou, 510515, China \and
Center for Advanced Medical Computing and Analysis (CAMCA), Massachusetts General Hospital and Harvard Medical School, Boston, 02114, USA\and
Department of Computer Science and Engineering, Lehigh University, Bethlehem, 18015, USA\and
School of Computing, The University of Georgia, Athens, 30602, USA}

\maketitle            

\renewcommand{\thefootnote}{\fnsymbol{footnote}}
\footnotetext[1]{These authors contributed equally to this paper.}
\footnotetext[4]{Corresponding authors.}

\begin{abstract}
Alzheimer's Disease (AD) is a neurodegenerative condition characterized by diverse progression rates among individuals, with changes in cortical thickness (CTh) closely linked to its progression. Accurately forecasting CTh trajectories can significantly enhance early diagnosis and intervention strategies, providing timely care. However, the longitudinal data essential for these studies often suffer from temporal sparsity and incompleteness, presenting substantial challenges in modeling the disease's progression accurately. Existing methods are limited, focusing primarily on datasets without missing entries or requiring predefined assumptions about CTh progression. To overcome these obstacles, we propose a conditional score-based diffusion model specifically designed to generate CTh trajectories with the given baseline information, such as age, sex, and initial diagnosis. Our conditional diffusion model utilizes all available data during the training phase to make predictions based solely on baseline information during inference without needing prior history about CTh progression. The prediction accuracy of the proposed CTh prediction pipeline using a conditional score-based model was compared for sub-groups consisting of cognitively normal, mild cognitive impairment, and AD subjects. The Bland-Altman analysis shows our diffusion-based prediction model has a near-zero bias with narrow 95\% confidential interval compared to the ground-truth CTh in 6-36 months. In addition, our conditional diffusion model has a stochastic generative nature, therefore, we demonstrated an uncertainty analysis of patient-specific CTh prediction through multiple realizations.

\end{abstract}

\section{Introduction}
Alzheimer's Disease (AD), as one of the most common neurodegenerative diseases \cite{yang2021deep}, has an insidious onset, with the progressive decline of cognitive and behavioral functions and irreversible brain atrophy. 
Typically, the full progression of AD, from a cognitively normal (CN) state, through mild cognitive impairment (MCI), to eventual AD, may span many years. The transition between these states—or the rate of deterioration—varies significantly among individuals and at different stages of the disease \cite{yi2023identifying}. 
Fortunately, the change of biomarkers can correspond to specific groups, providing a new insight for predicting future disease progression \cite{xu2022multi}. The relationships between the trajectory of biomarkers, such as changes in cortical thickness (CTh), and the gradual progression of AD raises the necessity for longitudinal data. 
However, longitudinal data in AD studies frequently suffer from sparsity (with significant gaps between consecutive visits) and incompleteness (where records for certain visits may be missing for various reasons), posing significant challenges for modeling disease progression.

According to \cite{schwarz2016large}, the reduction in CTh has been closely associated with the advancement of AD, making CTh a valuable biomarker for predicting tasks related to AD. Some literature has been proposed to model the progression of CTh. 
For example, P{\'e}rez-Millan et al. \cite{CT1} utilized support vector regression to fit the cortical thickness model and predict longitudinal visits using baseline data.
Marinescu et al. \cite{CT2} introduced a parametric model for tracking the progression of AD, designed to extrapolate long-period patterns of brain pathology from limited short-period longitudinal datasets. However, these existing methods either focus on the complete data or require prior knowledge of disease trajectories, which can be challenging to ascertain beforehand, especially for diseases with high variability. Recently, deep learning techniques have demonstrated encouraging outcomes in uncovering patterns within complex datasets. In particular, diffusion models, as a subset of generative models, have attracted considerable interest across various fields due to their remarkable capabilities\cite{yang2023diffusion}. Unlike deterministic models such as recurrent neural networks (RNN), diffusion models are adept at managing uncertainties thanks to their inherent stochastic mechanism. Furthermore, conditional diffusion models offer a versatile approach for generating targeted outputs under specific conditions. Nonetheless, the application of diffusion models in disease progression modeling remains sparse. 

In this paper, we propose a conditional score-based diffusion model to predict different trajectories of CTh. 
To our knowledge, our model represents the first attempt at longitudinal prediction of CTh based on the diffusion model. 
We employ the ``elucidating the design space of diffusion-based generative models" (EDM) framework as the backbone \cite{EDM}. Our model takes three conditions to guide the diffusion process: the clinical information (including diagnosis, sex, and age), the CTh of the previous visit, and the time interval between different scans. We evaluated our conditional score-based diffusion model using the Alzheimer's Disease Prediction Of Longitudinal Evolution (TADPOLE) challenge\footnote{https://tadpole.grand-challenge.org/} cohort. Population, sub-group, and patient-specific analysis demonstrated the effectiveness of our model in the longitudinal CTh prediction. 

\section{Method}

\subsection{Data Description}

In our study, we utilized TADPOLE Challenge cohort, a comprehensive dataset derived from the Alzheimer’s Disease Neuroimaging Initiative\footnote{https://adni.loni.usc.edu/}. Our analysis considered various variables, including diagnosis, age, gender, and the longitudinal CTh across 68 regions of interest (ROIs) according to the Desikan-Killiany cortical atlas \cite{DK_atlas}. The measurement of CTh was obtained through a sequential longitudinal processing pipeline in Freesurfer \cite{reuter2012within}, encompassing steps such as registration, normalization, skull stripping, segmentation, and parcellation. 

In our study, we successfully enrolled a total of 898 participants across five distinct time points: baseline (bl), 6 months (m06), 12 months (m12), 24 months (m24), and 36 months (m36). All enrolled participants had complete data at baseline, however, varying degrees of missing data were observed in subsequent follow-up visits. Therefore, we allocated 178 participants with complete data to the testing set and the remaining 720 participants to the training set. Initial diagnoses at baseline revealed that the training cohort included 187 AD, 324 MCI, and 209 CN. The testing cohort initially consisted of 100 MCI and 78 CN individuals. By the 36-month mark, the composition of the testing cohort had evolved to include 40 AD, 68 MCI, and 70 CN individuals.

\begin{figure}[t]
\centering
\includegraphics[width=0.87\textwidth]{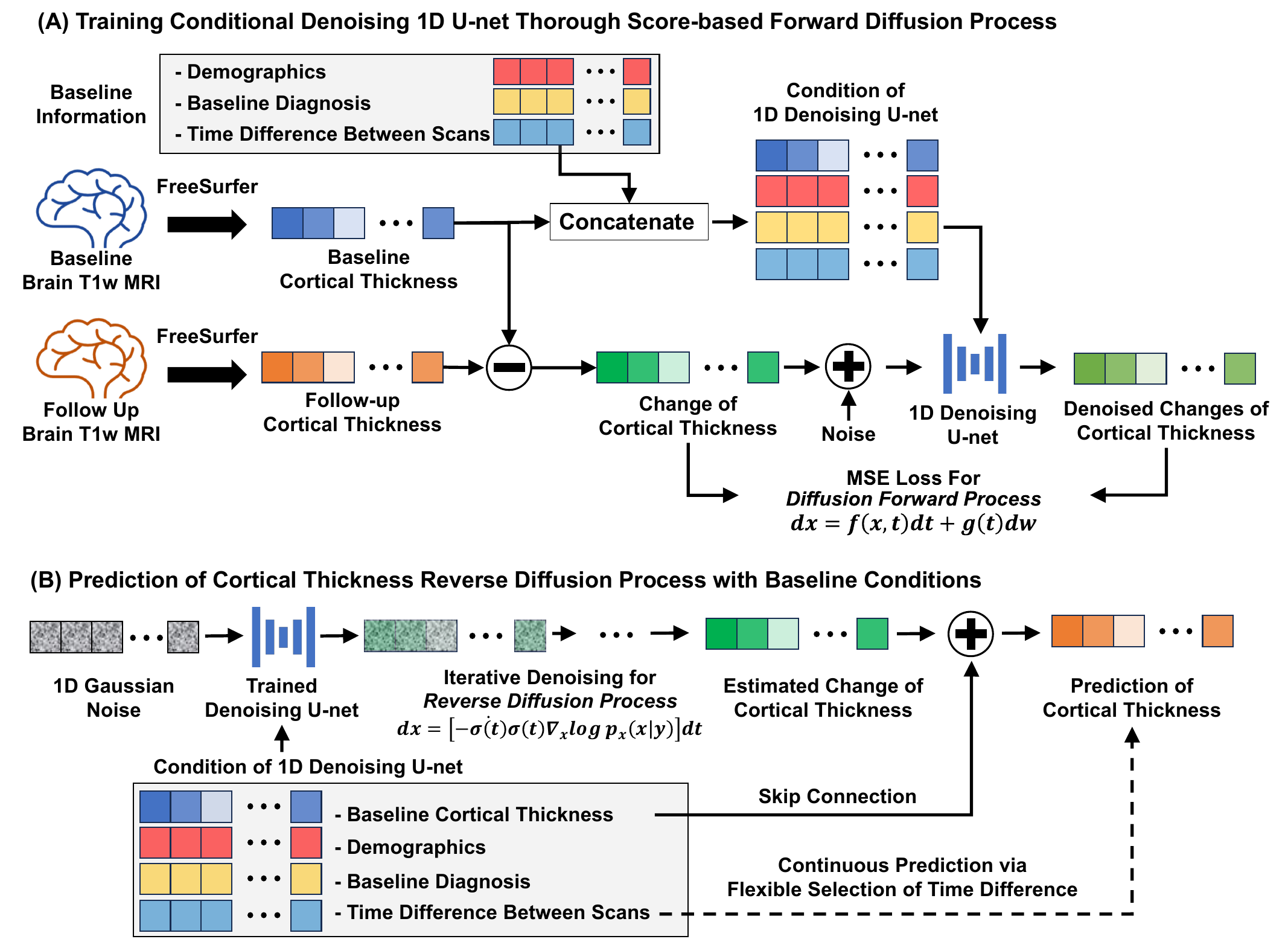}
\caption{\textbf{Overview of the score-based diffusion model framework for cortical thickness (CTh) prediction.} 
(A) Training of the conditional score-based diffusion model is achieved through a forward diffusion process integrating baseline information such as demographics, diagnosis, and inter-scan interval with CTh measurements. The baseline and follow-up T1-weighted MRIs are pre-processed to extract the CTh using FreeSurfer. The concatenated baseline information is then fed through the 1D denoising U-net to estimate the score function, which is used in the reverse diffusion process. The 1D denoising U-net is trained by MSE loss to estimate gradual changes in data distributions from baseline information to prediction of CTh changes. 
(B) The prediction stage employs a reverse diffusion process conditioned on baseline characteristics and CTh, utilizing a trained denoising U-net and iterative denoising to estimate changes in CTh, which can then be used to predict future CTh trajectories. Note that our framework supports continuous prediction through a flexible selection of time differences. } 
\label{fig1}
\end{figure}
\subsection{Conditional Score-based Diffusion Model for CTh Prediction}
Previous research has indicated a connection between the progression of AD and patterns of CTh \cite{hwang2016prediction,kim2024distinct}. Additionally, several clinical factors, such as age and sex, may affect different variations in CTh across specific brain regions \cite{gennatas2017age}. 
Thus, our diffusion model is conditioned by the baseline information provided (Fig. \ref{fig1}), such as the initial status of CTh, other available demographics, initial diagnosis, and the time difference between the baseline and the target prediction time. This represents the multifaceted nature of the progression of thinning of the cerebral cortex. 

The diffusion model consists of forward and reverse diffusion processes. During the forward diffusion process, the diffusion model gradually introduces Gaussian noise to the conditional data distribution $p(x|y)$, creating a sequence of progressively noisier samples as described by the following equation:
\begin{equation}
    \label{eq1}
    p(x_t|x_0, y) = \mathbf{N}(x_t; \mu_t x_0, \mu_t^2 \sigma_t^2 I),
\end{equation}
where $x$ is a residual of two CTh measurements at different time points, and $y$ is a given baseline condition, such as clinical characteristics and initial CTh. This process is a continuous transformation that transitions the data distribution to a prior distribution over time $t$ ranging from 0 to $T$, where the initial state $x(0)$ follows the data distribution $p_0$, and the final state $x(T)$ aligns with the prior distribution $p_T$. 

In the context of score-based generative models, this transformation is depicted by a stochastic differential equation (SDE):
\begin{equation}
    \label{eq2}
 dx = f(x,t)dt + g(t)dw,
\end{equation}
where $w$ represents Brownian motion, $f(\cdot,t)$ is the drift function, and $g(t)$ denotes the diffusion coefficient. 

A reverse-time SDE governs the reverse process to generate samples from the data distribution $x(0) \sim p_0$ starting from the prior distribution $x(T) \sim p_T$. The conditional reverse diffusion process can be formulated into an ordinary differential equation that offers a deterministic perspective of the diffusion process by eliminating the stochastic component (Brownian motion) \cite{EDM}:
\begin{equation}
    \label{eq3}
 dx = [-\dot{\sigma(t)}\sigma(t) \nabla_x \log p_t(x|y)]dt, 
 \end{equation}
where the score function $\nabla_x \log p_t(x|y)$ represents the gradient of the log-likelihood of the data distribution to the data itself. For a given data point $x$ and baseline condition $y$, the conditional score function guides this denoising process by indicating how to adjust the data at each step to make it more likely under the training data distribution. To effectively execute the reverse process, the score function, $\nabla_x \log p_t(x|y)$, which can be estimated using a neural network trained through score matching alongside a denoiser function $D(x;\sigma,y)$:
\begin{equation}
    \label{eq4}
 \nabla_x \log p_t(x|y) = (D(x;\sigma,y) - x) / \sigma^2. 
\end{equation}

We trained denoiser function $D(x;\sigma,y)$ via a mean squared error (MSE) loss:
\begin{equation}
    \label{eq5}
 L := \mathbf{E}_{p_{data}} \mathbf{E}_{n \sim \textit{N}(0, \sigma^2 I)} \| D(x+n; \sigma, y) - x \|^2_2.
\end{equation}
Once the model is trained, we use the trained model to sample and estimate the change of CTh in the reverse process using Equation (\ref{eq3}).

\subsection{Training and Implementation Details}
Our conditional score-based diffusion model is based on the EDM framework \cite{EDM}. We have modified the neural network architecture to a 1D Attention U-net, which takes the CTh values of ROIs as a 1-D tensor with a size of 68 and uses baseline information as input. We expanded the baseline information to the same size as the 1D CTh tensor and concatenated it in the channel dimension for network conditioning. The 1D Attention U-net was trained on a single NVIDIA A100 GPU with 40GB of memory and a batch size of 64 for 8192 epochs using Adam optimizer with a learning rate of 0.001. The number of function evaluations was 1000 for reverse sampling. Our source code and trained model will be available online. 

\section{Results}
\begin{table}[t]
    \centering
    \caption{Comparison MAE results (mean$\pm$SD) obtained by various methods on predicting longitudinal cortical thickness within the entire study cohort (All) and three subgroups (CN, MCI, and AD), respectively. }
    {
    \begin{tabular}{ccccc}
        \toprule
        Method & CN & MCI & AD & All \\
        \midrule
        cFSGL\cite{cFSGL} & 0.142$\pm$0.088 & 0.162$\pm$0.135 & 0.144$\pm$0.085 & 0.150$\pm$0.108 \\
        GRUD \cite{GRUD} & 0.109$\pm$0.016 & 0.121$\pm$0.035 & 0.124$\pm$0.022 & 0.117$\pm$0.026 \\
        LSTM-T \cite{LSTMT} & 0.114$\pm$0.015 & 0.124$\pm$0.031 & 0.125$\pm$0.021 & 0.120$\pm$0.024 \\
        Ours & \textbf{0.082$\pm$0.018} & \textbf{0.096$\pm$0.046} & \textbf{0.099$\pm$0.017} & \textbf{0.092$\pm$0.032}\\
        \bottomrule
    \end{tabular}}
    \label{table1}
\end{table}

\subsection{Comparison Results with Related Methods}
We compared the predictive results of the longitudinal CTh from our proposed method with the following models. Firstly, Zhou et al. \cite{cFSGL} introduced a convex fused sparse group lasso (cFSGL), which treats the prediction at each time point as a separate task and leverages baseline features to forecast longitudinal outcomes. Secondly, Che et al. \cite{GRUD} combined masking and time interval with gated recurrent unit (GRU) to capture the long-term temporal dependencies. This method employs a weighted integration of the most recent observation, the empirical mean, and a recurrent component for data imputation.  Lastly, Jung et al. \cite{LSTMT} used a deep recurrent network based on long short-term memory (LSTM) for personalized AD prediction via intrinsic temporal and multivariate relations.

We compared the mean absolute error (MAE) across all patients and all time points for all methods under comparison. The data presented in Table \ref{table1} suggest that non-linear models (including RNN-based models and our model) outperform the linear model (cFSGL), likely due to their ability to capture the more complex dynamics inherent in longitudinal changes.
Furthermore, our diffusion model surpasses all other models in terms of performance across the entire testing cohort and all three specific subgroups. This superiority may be attributed to incorporating the diffusion process within our model, enabling a more nuanced understanding and prediction of CTh trajectories. 
Lastly, predicting the CTh trajectory appears more challenging within the MCI and AD groups than in the CN group, as indicated by the MAE results. A plausible explanation could be that subjects with MCI and AD may undergo more obvious and diverse changes in CTh, leading to more heterogeneous developments among individuals during follow-up visits compared to those observed in CN subjects.

\subsection{Ablation Study}
\begin{table}[t]
    \centering
    \caption{MAE results (mean$\pm$SD) from the ablation study on predicting longitudinal cortical thickness, within the entire study cohort (All) and three subgroups (CN, MCI, and AD), respectively. U-net models with and without the attention mechanism are referred to as U-net (w/ a) and U-net (w/o a), respectively.}{
    \begin{tabular}{ccccc}
        \toprule
        Method & CN & MCI & AD & All \\
        \midrule
        U-net (w/o a) & 0.098$\pm$0.017 & 0.112$\pm$0.039 & 0.117$\pm$0.022 & 0.108$\pm$0.029 \\
        U-net (w/ a) & 0.084$\pm$0.025 & \textbf{0.092$\pm$0.039} & 0.116$\pm$0.029 & 0.094$\pm$0.034 \\
        Ours & \textbf{0.082$\pm$0.018} & 0.096$\pm$0.046 & \textbf{0.099$\pm$0.017} & \textbf{0.092$\pm$0.032}\\
        \bottomrule
    \end{tabular}}
    \label{table2}
\end{table}

To assess the impact of the diffusion process on predicting CTh, we introduced two variants of network architecture: U-net (w/o a) and U-net (w/ a). The distinguishing feature between the two is incorporating an attention module in U-net (w/ a), which is absent in U-net (w/o a). The U-net (w/o a) and U-net (w/ a) were trained using the identical setting and hyperparameters to the diffusion model, while training loss was given as an MSE loss for the prediction of CTh change in a supervised manner. Unlike our proposed diffusion-based model, these variants operate deterministically, predicting CTh changes. The findings in Table \ref{table2} demonstrate that the attention mechanism enhances predictive accuracy. While our diffusion model outperforms in most scenarios, it falls slightly behind within the MCI group compared with U-net (w/ a). However, a significant advantage of our model is its ability to quantify prediction uncertainty, which was not available in the deterministic variants. This capability to predict with uncertainty offers more dependable support for clinical decision-making processes.

\subsection{Group-wise Analysis of Our Method}
\begin{figure}[t]
\centering
\includegraphics[width=0.94\textwidth]{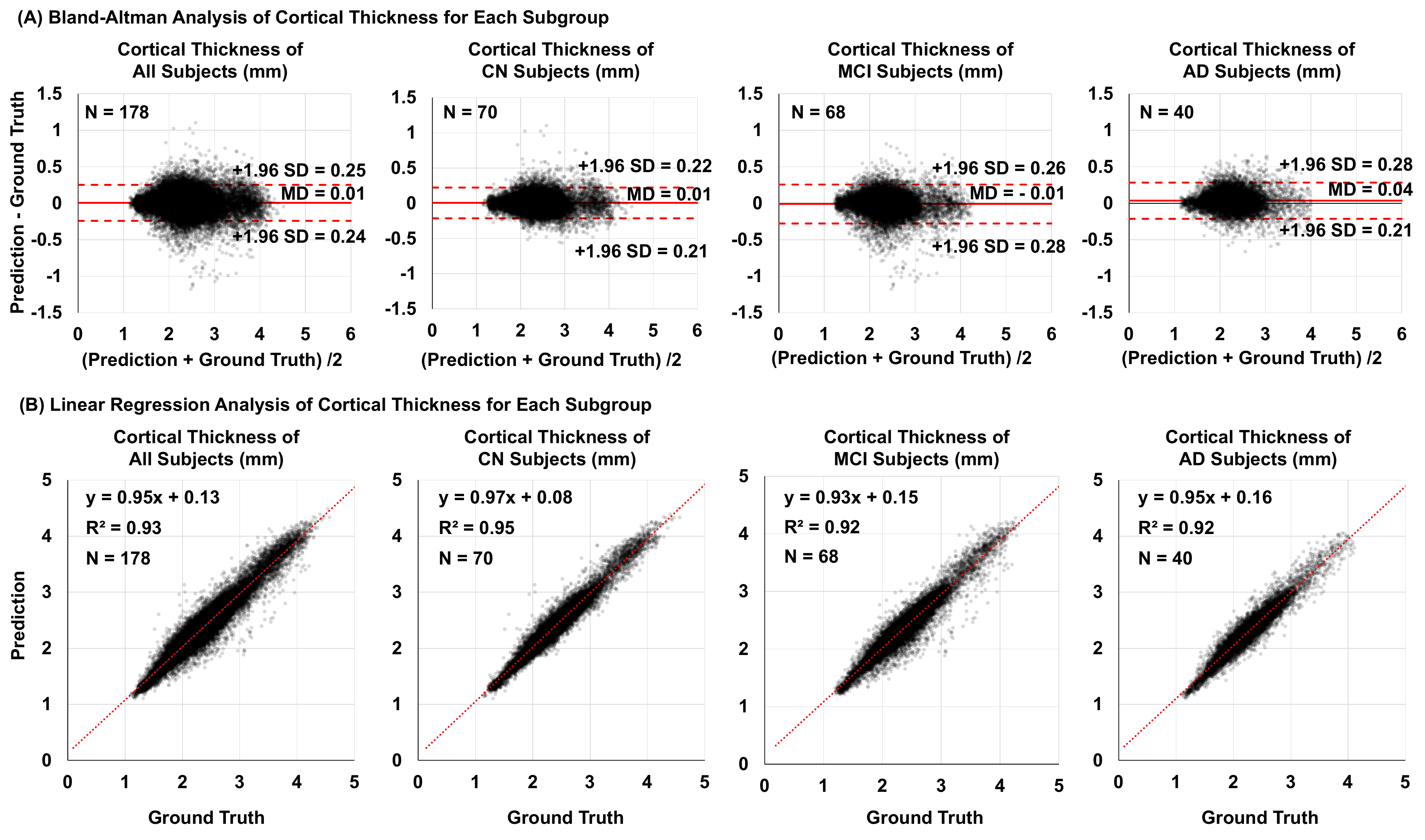}
\caption{\textbf{Bland-Altman analysis and correlation of predicted and measured cortical thickness across different subgroups.} (A) From left to right, Bland-Altman plots show the agreement between predicted and actual cortical thickness for all subjects (N=178), CN subjects (N=70), MCI subjects (N=68), and AD subjects (N=40), respectively, with mean differences (MD) near zero predictive bias. (B) Scatter plots with linear regression analysis demonstrate strong correlations (R$^2>$ 0.9) between predicted and actual cortical thickness across all subjects and all subgroups.} 
\label{fig2}
\end{figure}

To delve deeper into the predictive accuracy of our proposed model, we performed both Bland-Altman analysis and linear regression analysis to compare our predictions with the ground truth across all subjects and subgroups, as depicted in Figure \ref{fig2}. 
In the Bland-Altman analysis, the proximity of all mean differences (MD) lines to zero suggests negligible predictive bias, with most differences residing within the mean $\pm$ 1.96 SD range, as shown in Figure \ref{fig2}(A). There's a remarkable concordance between the predicted CTh and the actual measurements across all examined groups. Furthermore, the linear regression analysis, as illustrated in Figure \ref{fig2}(B), reveals a strong positive correlation between the predicted CTh and ground truth across all groups under comparison, with both the correlation coefficient (R$^2$) and the slope above 0.9.

\subsection{Patient-specific Analysis of Our Method}
\begin{figure}[t]
\centering
\includegraphics[width=0.9\textwidth]{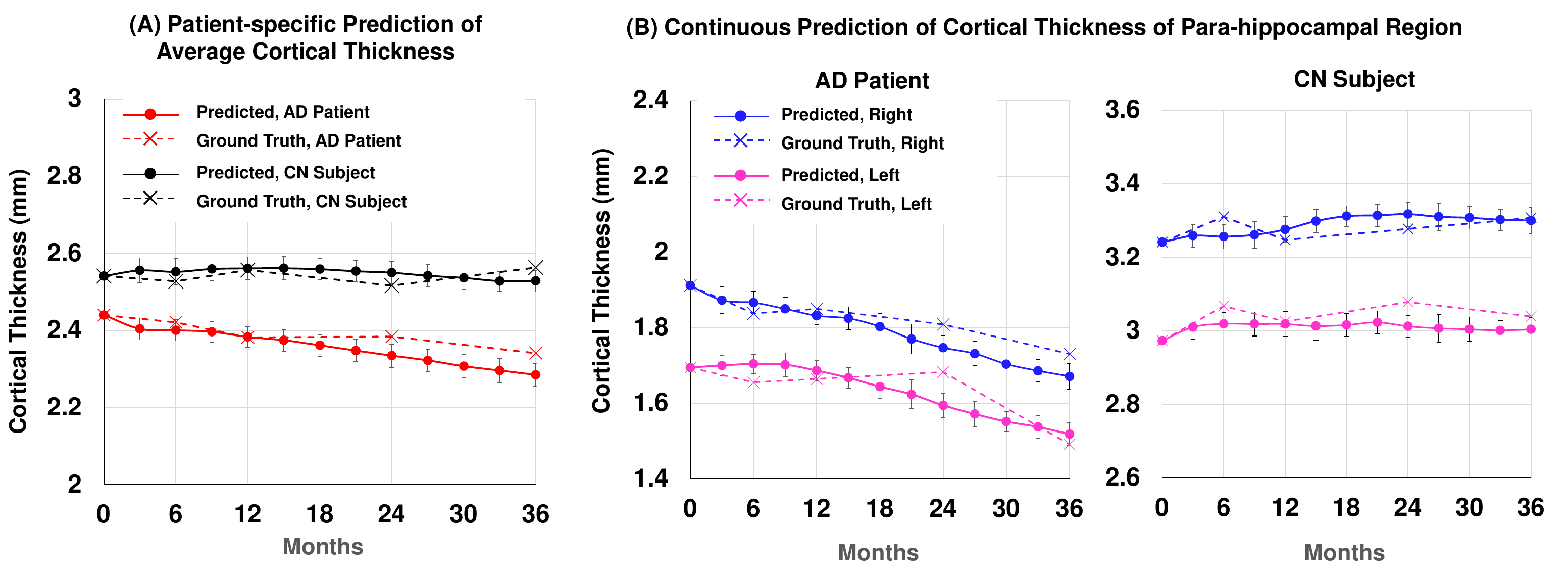}
\caption{\textbf{Longitudinal cortical thickness predictions in AD and CN exemplar subjects. }(A) Continuous prediction versus actual sparse measurement of average cortical thickness in a 60-year-old male AD patient and 71-year-old male CN subject over 36 months. Predicted values for AD (red) demonstrate more obvious cortical thinning, while data for CN subjects (black) exhibit relative stability. (B) Separate analyses of the left and right para-hippocampal regions show a similar pattern, with the AD patient exhibiting a more pronounced decline in cortical thickness than the CN subject. The error bar of predicted values demonstrates the uncertainty of the model through multiple realizations.} 
\label{fig3}
\end{figure}

Our framework not only facilitates continuous prediction but also enables the analysis of uncertainty, which has significant value in clinical settings. Consequently, we selected two representative subjects for illustrative purposes: one 60-year-old male from the AD group and another 71-year-old male from the CN group in Figure \ref{fig3}. Although the two samples have similar starting points at baseline, as time progresses, the gap in average CTh between the AD and CN samples widens, indicating a more pronounced trend of CTh thinning in the AD sample compared to the CN sample. Figure \ref{fig3}(B) further illustrates the predictive CTh outcomes for the para-hippocampal region in both the left and right hemispheres. Despite the disparities in CTh values between the two hemispheres, a consistent pattern emerges—the AD subject demonstrates a more noticeable decrease in CTh values than the CN subject. The uncertainty of the model in the patient-specific analysis is shown as an error bar, which is calculated through multiple stochastic realizations.

\section{Conclusion}
In this paper, we proposed a conditional score-based diffusion model to forecast CTh progression solely based on the baseline information. Our model was evaluated at different scales, including population, sub-group, and patient-specific. The analyses demonstrate the potential of our method in the prediction of longitudinal CTh, which may benefit the diagnosis and intervention strategies in the preclinical stages of AD. Our diffusion model-based method can not only conduct continuous prediction of CTh values flexibly but also provide prediction uncertainty, which holds significant value in clinical settings. 

\bibliographystyle{splncs04}
\bibliography{reference}

\end{document}